\documentclass{cernphprep}

\usepackage{times}
\usepackage{graphicx}
\usepackage{bm}

\begin{document}

\begin{titlepage}

\EXPnumber{DIRAC/PS212}
\PHnumber{2014--30}
\PHdate{\today}

\title{First $\pi K$ atom lifetime and $\pi K$ scattering length measurements}

\begin{Authlist}

B.~Adeva\Iref{s}, 
L.~Afanasyev\Iref{d}, 
Y.~Allkofer\Iref{zu}, 
C.~Amsler\Iref{be}, 
A.~Anania\Iref{im},
S.~Aogaki\Iref{b},
A.~Benelli\Iref{d}, 
V.~Brekhovskikh\Iref{p},  
T.~Cechak\Iref{cz}, 
M.~Chiba\Iref{jt}, 
P.~Chliapnikov\Iref{p}, 
C.~Ciocarlan\Iref{b}, 
S.~Constantinescu\Iref{b}, 
P.~Doskarova\Iref{cz}, 
D.~Drijard\Iref{c},
A.~Dudarev\Iref{d},
M.~Duma\Iref{b},  
D.~Dumitriu\Iref{b}, 
D.~Fluerasu\Iref{b}, 
A.~Gorin\Iref{p}, 
O.~Gorchakov\Iref{d},
K.~Gritsay\Iref{d}, 
C.~Guaraldo\Iref{if}, 
M.~Gugiu\Iref{b}, 
M.~Hansroul\Iref{c}, 
Z.~Hons\Iref{czr}, 
S.~Horikawa\Iref{zu},
Y.~Iwashita\Iref{jk},
V.~Karpukhin\Iref{d}, 
J.~Kluson\Iref{cz}, 
M.~Kobayashi\Iref{k}, 
V.~Kruglov\Iref{d}, 
L.~Kruglova\Iref{d}, 
A.~Kulikov\Iref{d}, 
E.~Kulish\Iref{d},
A.~Kuptsov\Iref{d}, 
A.~Lamberto\Iref{im}, 
A.~Lanaro\Iref{u},
R.~Lednicky\Iref{cza}, 
C.~Mari\~nas\Iref{s},
J.~Martincik\Iref{cz},
L.~Nemenov\IIref{d}{c},
M.~Nikitin\Iref{d}, 
K.~Okada\Iref{jks}, 
V.~Olchevskii\Iref{d}, 
M.~Pentia\Iref{b}, 
A.~Penzo\Iref{it}, 
M.~Plo\Iref{s},
T.~Ponta\Iref{b},
P.~Prusa\Iref{cz},  
G.~Rappazzo\Iref{im}, 
A.~Romero Vidal\Iref{if},
A.~Ryazantsev\Iref{p},
V.~Rykalin\Iref{p}, 
J.~Schacher\IAref{be}{*},
A.~Sidorov\Iref{p}, 
J.~Smolik\Iref{cz}, 
S.~Sugimoto\Iref{k},
F.~Takeutchi\Iref{jks}, 
L.~Tauscher\Iref{ba},
T.~Trojek\Iref{cz}, 
S.~Trusov\Iref{m}, 
T.~Urban\Iref{cz},
T.~Vrba\Iref{cz},
V.~Yazkov\Iref{m}, 
Y.~Yoshimura\Iref{k}, 
M.~Zhabitsky\Iref{d}, 
P.~Zrelov\Iref{d} 

\end{Authlist}

\Instfoot{s}{Santiago de Compostela University, Spain}
\Instfoot{d}{JINR Dubna, Russia}
\Instfoot{zu}{Zurich University, Switzerland}
\Instfoot{be}{
Albert Einstein Center for Fundamental Physics, 
Laboratory of High Energy Physics, Bern, Switzerland
}
\Instfoot{im}{INFN, Sezione di Trieste and Messina University, Messina, Italy}
\Instfoot{b}{
IFIN-HH, National Institute for Physics and Nuclear Engineering, 
Bucharest, Romania
}
\Instfoot{p}{IHEP Protvino, Russia}
\Instfoot{cz}{Czech Technical University in Prague, Czech Republic}
\Instfoot{jt}{Tokyo Metropolitan University, Japan}
\Instfoot{c}{CERN, Geneva, Switzerland}
\Instfoot{if}{INFN, Laboratori Nazionali di Frascati, Frascati, Italy}
\Instfoot{czr}{Nuclear Physics Institute ASCR, Rez, Czech Republic}
\Instfoot{jk}{Kyoto University, Kyoto, Japan}
\Instfoot{k}{KEK, Tsukuba, Japan}
\Instfoot{u}{University of Wisconsin, Madison, USA} 
\Instfoot{cza}{Institute of Physics ASCR, Prague, Czech Republic}
\Instfoot{jks}{Kyoto Sangyo University, Kyoto, Japan}
\Instfoot{it}{INFN, Sezione di Trieste, Trieste, Italy}
\Instfoot{ba}{Basel University, Switzerland}
\Instfoot{m}{
Skobeltsin Institute for Nuclear Physics of Moscow State University, 
Moscow, Russia
}

\Anotfoot{*}{Corresponding author} 

\Collaboration{DIRAC Collaboration}
\ShortAuthor{DIRAC Collaboration}

\newpage

\begin{abstract}
The results of a search for hydrogen-like atoms 
consisting of $\pi^{\mp}K^{\pm}$ mesons are presented. 
Evidence for $\pi K$~atom production by 24~GeV/c~protons 
from CERN PS interacting with a nickel target has been 
seen in terms of characteristic $\pi K$~pairs from 
their breakup in the same target ($178 \pm 49$) and  
from Coulomb final state interaction ($653 \pm 42$). 
Using these results the analysis yields 
a first value for the $\pi K$ atom lifetime of 
$\tau=(2.5_{-1.8}^{+3.0})$~fs 
and a first model-independent measurement of 
the S-wave isospin-odd $\pi K$~scattering length 
$\left|a_0^-\right|=\frac{1}{3}\left|a_{1/2}-a_{3/2}\right|=
\left( 0.11_{-0.04}^{+0.09} \right)M_{\pi}^{-1}$ 
($a_I$ for isospin~$I$).
\end{abstract}
\vspace{2cm}
\Submitted{(To be submitted to Physics Letters B)}
\end{titlepage}

\section{Introduction}
\label{sec:intro}

In order to understand Quantum Chromodynamics (QCD) 
in the confinement region, low-energy QCD and 
specifically Chiral Perturbation Theory (ChPT) 
\cite{WEIN66,GASS85,MOUS00,COLA01} 
has to be explored and tested experimentally.
Pion-pion interaction at low energy is the simplest 
hadron-hadron process. The observation of dimesonic 
$\pi^+\pi^-$~atoms has been reported in \cite{AFAN94} 
and a measurement of their lifetime in \cite{ADEV05,ADEV11}. 

A measurement of the $\pi K$ atom\footnote{The term $\pi K$ atom or 
$A_{K \pi}$ refers to $\pi^- K^+$ and $\pi^+ K^-$ atoms.}  
lifetime provides a direct determination of an S-wave 
$\pi K$ scattering length difference \cite{BILE69}. This atom is 
an electromagnetically bound $\pi K$ state with a Bohr radius of  
$a_{B}$~=~249~fm and a ground state binding energy of $E_{B}$~=~2.9~keV. 
It decays predominantly\footnote{Further decay channels with photons 
and~~$\mathrm{e}^+\mathrm{e}^-$ pairs are suppressed at $\mathcal{O}(10^{-3})$.} 
by strong interaction into two neutral mesons 
$\pi^0 K^0$ or $\pi^0 \overline{K^0}$.  
The atom decay width~$\Gamma_{\pi K}$ in the ground state (1S)   
is given by the relation \cite{BILE69,SCHW04}:
\begin{equation}
  \label{eq:julia0}
   \Gamma_{\pi K} = \frac{1}{\tau} \simeq 
   \Gamma(A_{K \pi} \to \pi^0 K^0 \; or \; \pi^0 \overline{K^0}) = 
   8 \; \alpha^3 \; \mu^2 \; p^* \; (a_{0}^-)^2 \; (1+\delta_K).
\end{equation}

The S-wave isospin-odd $\pi K$ scattering length 
$a_{0}^-=\frac{1}{3}(a_{1/2}-a_{3/2})$,  
$a_I$~for isospin $I$, is defined in pure QCD for 
quark masses $m_u=m_d$~, 
$\alpha$ is the fine structure constant, 
$\mu$~=~109~$\rm{MeV/c^2}$ the reduced mass of 
the $\pi^{\mp}K^{\pm}$ system,  
$p^*$~=~11.8~MeV/c the outgoing $\pi^0$ or $K^0$ ($\overline{K^0}$) 
momentum in the $\pi K$ atom system, 
and $\delta_K$ accounts for corrections, due to 
isospin breaking, at order $\alpha$ and quark mass difference  
($m_u - m_d$) \cite{SCHW04}.
 
There is a remarkable evolution from 1966 to 2004 in $a_{0}^-$ 
calculation in the framework of SU(3)~ChPT and dispersion analysis:
\begin{equation}
  \label{eq:chpt}   
   M_{\pi} a_{0}^- = 
   0.071~(CA)~\rightarrow~
   0.0793 \pm 0.0006~(1l)~\rightarrow~
   0.089~(2l)~\rightarrow~
   0.090 \pm 0.005~(dis). 
\end{equation}
$CA$ denotes the current algebra value \cite{WEIN66}, $1l$ 
the prediction in SU(3)~ChPT at the 1-$loop$ level \cite{BERN91,KUBI02}, 
$2l$ correspondingly at 2-$loop$ \cite{BIJN04} and $dis$ 
the result of the dispersion analysis using 
Roy-Steiner equations \cite{BUET04} ($M_{\pi}$ is charged pion mass). 
Results from ongoing lattice simulations of 
$\pi K$ scattering \cite{LANG12} are expected in the near future. 

Inserting in (\ref{eq:julia0}) 
$M_{\pi} a_{0}^- = 0.090 \pm 0.005$ and $\delta_K = 0.040 \pm 0.022$ 
\cite{SCHW04}  
one predicts for the $\pi K$ atom lifetime 
\begin{equation}
  \label{eq:julia1}\
   \tau \simeq (3.5 \pm 0.4)\cdot10^{-15}~\text{s}. 
\end{equation}
This paper describes the first experimental measurement of $\tau$.

A method for producing and observing hadronic atoms  
has been developed \cite{NEME85} and successfully applied 
to $\pi^+ \pi^-$ atoms \cite{AFAN94,ADEV05,ADEV11}. 
The production yield of $\pi K$ atoms in proton-nucleus collisions 
has been calculated for different proton energies  
and atom emission angles \cite{GORC00}. In the DIRAC experiment 
relativistic dimesonic bound states, formed by Coulomb final state 
interaction, propagate inside a target and  
can break up (section~\ref{sec:Atom_int}).  
Particle pairs from breakup, called ``atomic pairs'' 
(atomic pair in Fig.~\ref{Fig3_1}), are 
characterized by small relative momenta, $Q <$ 3~MeV/c,  
in the centre-of-mass (c.m.) system of the pair. 
Here, $Q$ stands for the experimental c.m. relative momentum, 
smeared by multiple scattering in the target and other materials 
and by reconstruction uncertainties. Later, in the context of 
particle pair production, the original c.m. relative momentum 
$q$ will be used.

The results of the first $\pi K$ atom investigation have been    
published by DIRAC in 2008 \cite{ALLK08,ADEV09}: 
$\pi^- K^+$ and $\pi^+ K^-$ pairs are produced 
in a 26 $\mu m$ thick Pt target.   
An enhancement of $\pi K$ pairs at 
low relative momentum is observed, corresponding to  
$173 \pm 54$ $\pi K$ atomic pairs. 
The measured ratio of observed number of atomic pairs to  
number of produced atoms, the so-called breakup probability, 
allows to derive a lower limit on the $\pi K$ atom 
lifetime of $\tau > 0.8\cdot10^{-15}$~s (90\%~CL). 
For a real lifetime measurement a target material 
like Ni should be used because of its breakup probability 
rapidly rising with lifetime around $3.5\cdot10^{-15}$~s. 
 
Compared to the previous results \cite{ADEV09}, 
we present the analysis of a larger data sample collected from 
a Ni target by the DIRAC setup. By including information from 
detectors upstream of the spectrometer magnet the resolution 
in $Q$ is improved.

\section{Experimental setup}

\begin{figure}[ht]
\begin{center}
\includegraphics[width=100mm]{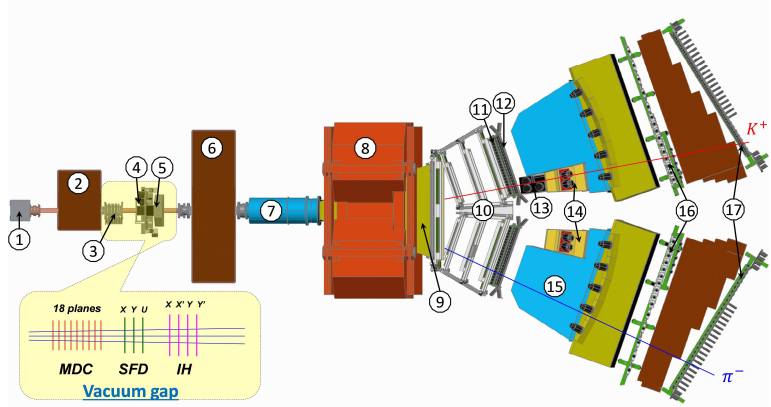}
\caption{ General view of the DIRAC setup:  
1 -- target station;
2 -- first shielding;
3 -- microdrift chambers;
4 -- scintillating fiber detector; 
5 -- ionisation hodoscope; 
6 -- second shielding; 
7 -- vacuum tube; 
8 -- spectrometer magnet; 
9 -- vacuum chamber; 
10 -- drift chambers; 
11 -- vertical hodoscope; 
12 -- horizontal hodoscope; 
13 -- aerogel Cherenkov; 
14 -- heavy gas Cherenkov; 
15 -- nitrogen Cherenkov; 
16 -- preshower; 
17 -- muon detector.}
\label{Fig2_1}
\end{center}
\end{figure}

The apparatus sketched in Fig.~\ref{Fig2_1}  
detects and identifies $\pi^+ \pi^-$, $\pi^- K^+$ and 
$\pi^+ K^-$ pairs with small $Q$. The structure of 
these pairs after the magnet is approximately 
symmetric for $\pi^+ \pi^-$ and asymmetric 
for $\pi K$. Originating from a bound system 
these particles travel with the same velocity, and 
therefore for $\pi K$ the kaon momentum is by a factor of 
about $\frac{M_{K}}{M_{\pi}}=3.5$ larger than the 
pion momentum ($M_{K}$ is charged kaon mass). 
The 2-arm magnetic spectrometer as presented 
is optimized for simultaneous detection of 
these pairs \cite{note0505,note0523}.

The 24~GeV/c primary proton beam from the CERN PS hits  
pure (99.98\%) Ni targets with thicknesses of ($98 \pm 1$) $\mu$m (Ni-1)  
in 2008 and ($108 \pm 1$) $\mu$m (Ni-2) in 2009 and 2010.
The radiation thickness of the 
98 (108) $\mu$m Ni target amounts to $6.7 \cdot 10^{-3}$ 
$(7.4 \cdot 10^{-3})$ $X_{0}$ (radiation length), which is optimal    
for the lifetime measurement. The nuclear interaction probability for 
98 (108) $\mu$m Ni is $6.4 \cdot 10^{-4}$ $(7.1 \cdot 10^{-4})$.

After the target station primary protons run forward to the beam dump, 
and the secondary channel with the whole setup is vertically inclined 
relative to the proton beam by $5.7^\circ$ upward.  Secondary 
particles are confined by the rectangular beam collimator inside of 
the second steel shielding wall, and the angular divergence in  
the horizontal (X) and vertical (Y) planes is $\pm 1^\circ$ and 
the solid angle $\Omega = 1.2 \cdot 10^{-3}$~sr. 
With a spill duration of 450~ms the beam intensity has been  
(10.5 -- 12) $\cdot10^{10}$ protons/spill and, correspondingly,   
the single counting rate in one plane of the ionisation 
hodoscope (IH) (5 -- 6) $\cdot 10^6$ particles/spill. 
Secondary particles propagate mainly in vacuum up 
to the Al foil with a thickness of 
0.68~mm $(7.6 \cdot 10^{-3} X_{0})$   
at the exit of the vacuum chamber, which is located between the  
poles of the dipole magnet ($B_{max}$ = 1.65~T and $BL$ = 2.2~Tm).

In the vacuum gap 18 planes of the MicroDrift Chambers (MDC) 
and 3 planes (X, Y, U) of the Scintillating Fiber Detector (SFD) 
have been installed to measure the particle coordinates   
($\sigma_{SFDx} = \sigma_{SFDy} = 60~\mu$m, $\sigma_{SFDu} = 120~\mu$m)
and the particle time   
($\sigma_{tSFDx} = 380$~ps, $\sigma_{tSFDy} = \sigma_{tSFDu} = 520$~ps).
The four IH planes serve to identify 
unresolved double tracks (signal only from one SFD column). 
The total matter radiation thickness between target and vacuum chamber 
amounts to $5.6 \cdot 10^{-2} X_{0}$.

Each spectrometer arm is equipped with the following subdetectors \cite{DIRA13}: 
drift chambers (DC) to measure particle coordinates with 
$\approx$85 $\mu$m precision; 
vertical hodoscope (VH) to measure time with 110 ps accuracy 
for particle identification via time-of-flight determination; 
horizontal hodoscope (HH) to select in the two arms particles 
with vertical distances less than 75 mm ($Q_{Y}$ less than 15 MeV/c); 
aerogel Cherenkov counter (ChA) to distinguish kaons from protons; 
heavy gas ($C_{4}F_{10}$) Cherenkov counter (ChF) to distinguish 
pions from kaons; 
nitrogen Cherenkov (ChN) and preshower (PSh) detector  
to identify $\mathrm{e}^+\mathrm{e}^-$ pairs; 
iron absorber; 
two-layer muon scintillation counter (Mu) to identify muons.
In the ``negative'' arm no aerogel counter has been installed, 
because the number of antiprotons is small compared to $K^{-}$.

Pairs of oppositely charged particles, time-correlated (prompt pairs)  
and accidentals in the time interval $\pm 20$~ns, are selected by 
requiring a 2-arm coincidence (ChN in anticoincidence) with 
a coplanarity restriction (HH) in the first-level trigger. 
The second-level trigger selects events with at least one track 
in each arm by exploiting DC-wire information (track finder).
Using track information the online trigger selects $\pi \pi$ and 
$\pi K$ pairs with $|Q_X| < 12~\rm{MeV/c}$ and 
$|Q_L| < 30~\rm{MeV/c}$~\footnote{The 
transverse $(Q_T=\sqrt{Q^2_X + Q^2_Y}~)$ and 
longitudinal ($Q_L$) components of $\vec Q$ are defined with 
respect to the direction of the total laboratory pair momentum.}.
The trigger efficiency is $\approx$ 98\% for pairs with 
$|Q_X| < 6~\rm{MeV/c}$, $|Q_Y| < 4~\rm{MeV/c}$ and $|Q_L| < 28~\rm{MeV/c}$.
For spectrometer calibration $\pi^- p$ ($\pi^+ \bar{p}$) pairs from 
$\Lambda$ ($\bar{\Lambda}$) decay have been investigated, 
and $\mathrm{e}^+\mathrm{e}^-$ pairs for general detector calibration.

\section{Production of bound and free and $\pi^- K^+$ and $\pi^+ K^-$ pairs}
\label{freebound}

\begin{figure}[ht]
\begin{center}
\includegraphics[width=75mm]{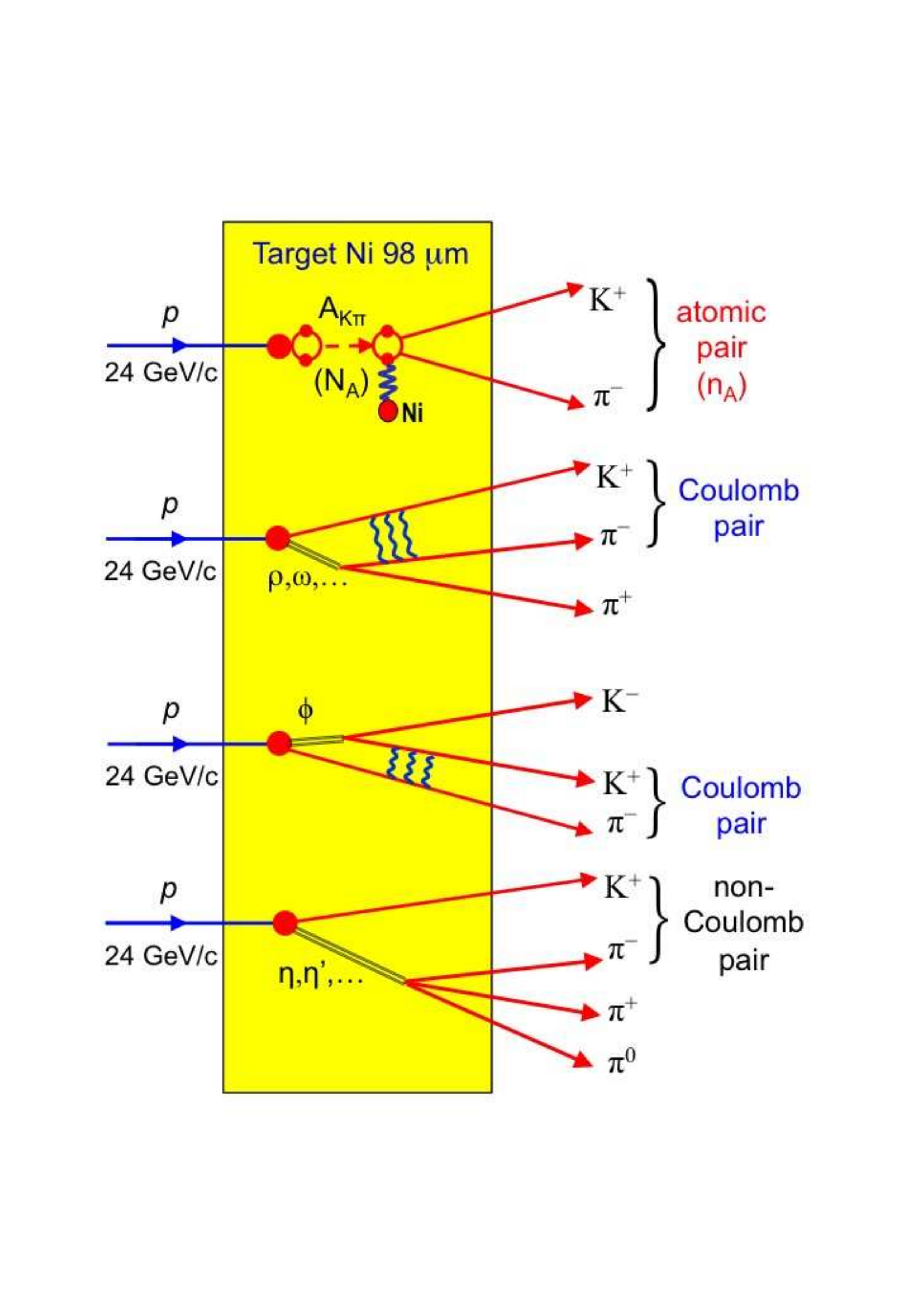}
\caption{Inclusive $\pi K$ production in  
24~GeV/c p-Ni interaction: 
p + Ni $\to$ $\pi^{\mp} K^{\pm}$ + X. 
The ionisation or breakup of $\pi K$ atoms, $A_{K \pi}$, leads to 
so-called atomic pairs. (More details, see text)}
\label{Fig3_1}
\end{center}
\end{figure}

Prompt $\pi^{\mp}K^{\pm}$ pairs from 
proton-nucleus collisions are produced either directly 
or originate from 
short-lived (e.g. $\Delta$, $\rho$), 
medium-lived (e.g. $\omega$, $\phi$) or 
long-lived (e.g. $\eta'$, $\eta$) sources. 
Pion-kaon pairs produced directly, from short- and medium-lived 
sources undergo Coulomb final state interaction 
(Coulomb pair in Fig.~\ref{Fig3_1}) and so 
may form bound states. Pairs from long-lived sources 
are practically not affected by Coulomb interaction 
(non-Coulomb pair in Fig.~\ref{Fig3_1}).
The accidental pairs are produced in different 
proton-nucleus interactions.

The cross section of $\pi K$ atom production 
is given by the expression \cite{NEME85}: 
\begin{equation}\label{eq:prod}
\frac{{\rm d}\sigma^{n}_A}{{\rm d}\vec p_A}=(2\pi)^3\frac{E_A}{M_A}
\left.\frac{{\rm d}^2\sigma^0_s}{{\rm d}\vec p_K {\rm d}\vec p_\pi}\right |_{
\frac{\vec p_K}{M_{K}} \approx \frac{\vec p_\pi}{M_{\pi}}}
\hspace{-1mm} \cdot \left|\psi_{n}(0)\right|^2 = 
(2\pi)^3\frac{E_A}{M_A}
\frac{1}{\pi a_B^3 n^3}
\left.\frac{{\rm d}^2\sigma^0_s}{{\rm d}\vec p_K {\rm d}\vec p_\pi}\right |_{
\frac{\vec p_K}{M_{K}} \approx \frac{\vec p_\pi}{M_{\pi}}} \:,
\end{equation}
where $\vec p_{A}$, $E_{A}$ and $M_{A}$ are the momentum, 
total energy and mass of the $\pi K$ atom in the 
laboratory (lab) system, respectively, and $\vec p_K$ and 
$\vec p_\pi$ the momenta of the charged kaon and pion 
with equal velocities. Therefore, these momenta obey 
in good approximation the relations 
$\vec p_{K}=\frac{M_{K}}{M_{A}} \vec p_{A}$ and 
$\vec p_{\pi}=\frac{M_{\pi}}{M_{A}} \vec p_{A}$. 
The inclusive production cross section of 
$\pi K$ pairs from short-lived sources without 
final state interaction (FSI) is denoted by $\sigma_s^0$, 
and $\psi_{n}(0)$ is the $S$-state atomic wave function  
at the origin with principal quantum number $n$.  
According to (\ref{eq:prod}) $\pi K$ atoms are only produced in 
$S$-states with probabilities $W_n~=~\frac{W_1}{n^3}$:      
$W_1~=~83.2\%$, $W_2~=~10.4\%$, $W_3~=~3.1\%$, $\dots$ , 
$W_{n>3}~=~3.3\%$.

In complete analogy, the production of free $\pi^{\mp}K^{\pm}$ pairs 
from short- and medium-lived sources,
i.e. Coulomb pairs, is described in 
the pointlike production approximation in dependence of 
relative momentum $q$ (section~\ref{sec:intro}) by 
\begin{equation}\label{eq:cross_sect_C}
\frac{{\rm d}^2\sigma_C}{{\rm d}\vec p_K {\rm d}\vec p_\pi} =
\frac{{\rm d}^2\sigma^0_s}{{\rm d}\vec p_K {\rm d}\vec p_\pi}
\hspace{-2mm} \cdot A_C(q) 
\quad \mbox{with} \quad
A_C(q) = \frac{4\pi \mu \alpha/q}
{1-\exp\left(-4\pi \mu \alpha/q\right) } \;.
\end{equation}
The Coulomb enhancement function $A_C(q)$ is the well-known 
Sommerfeld-Gamov-Sakharov factor \cite{SOMM21,GAMO28,SAKH91}.

The relative yield between atoms and Coulomb pairs \cite{AFAN99}
is given by the ratio of  
(\ref{eq:prod}) to (\ref{eq:cross_sect_C}). 
The total number $N_A$ of produced $\pi K$ atoms  
is determined by the model-independent relation 
\begin{equation}\label{eq:number_A}
N_A = k(q_0) N_C(q \le q_0) 
\quad \mbox{with} \quad
k(q_0 = 3.12~\rm{MeV/c}) = 0.615 \:,
\end{equation}
where $N_C(q \le q_0)$ is the number of Coulomb pairs 
with relative momenta $q \le q_0$ and  $k(q_0)$ 
a known function of $q_0$. 
By using the Monte Carlo (MC) technique, one gets the same 
relationship as in (\ref{eq:number_A}), but this time 
in terms of the experimental relative momentum $Q$.

So far the pair production is assumed to be pointlike. 
In order to check for finite size effects due to the presence 
of medium-lived particles ($\omega$, $\phi$), a study of  
non-pointlike particle pair sources has been performed~\cite{LEDN08}.  
Due to the large value of the Bohr radius, $a_B = 249$ fm, 
the pointlike treatment of the Coulomb $\pi K$ FSI is valid 
for directly produced pairs as well as for pairs from 
short-lived resonances. 
For $\pi$ and $K$ from medium-lived sources, 
corrections at the percent level have been applied to  
the production cross sections \cite{LEDN08}. 
Strong final state elastic and inelastic $\pi K$ interactions 
are negligible.

\section{Interaction of $\pi K$ and $\pi \pi$ atoms with matter}
\label{sec:Atom_int}

While propagating through the target material,
relativistic $\pi K$~atoms can get excited or even ionised.  
The ionisation or breakup process competes  
with $\pi K$~atom annihilation. 
The breakup probability $P_{\text{br}}$ 
as a function of the atom lifetime~$\tau$, 
atom momentum~$p_A$, target material and thickness 
has been extensively studied in the pionium case. 
To guarantee knowledge of $P_{\text{br}}(\tau,p_A)$ at 
the 1\%~level, one has to take into account 
a series of projectile collisions 
with matter atoms along the path in the target, 
leading to transitions between 
various bound states or to breakup. For $\pi \pi$~atoms 
the resulting system of equations is solved exactly by 
eigendecomposition of the corresponding matrix~\cite{Afan96,Zhab08}  
or by MC simulations~\cite{Santa03}. The same approach 
can be applied for $\pi K$~atoms.

\begin{figure}[!ht]
\begin{center}
\includegraphics[width=75mm]{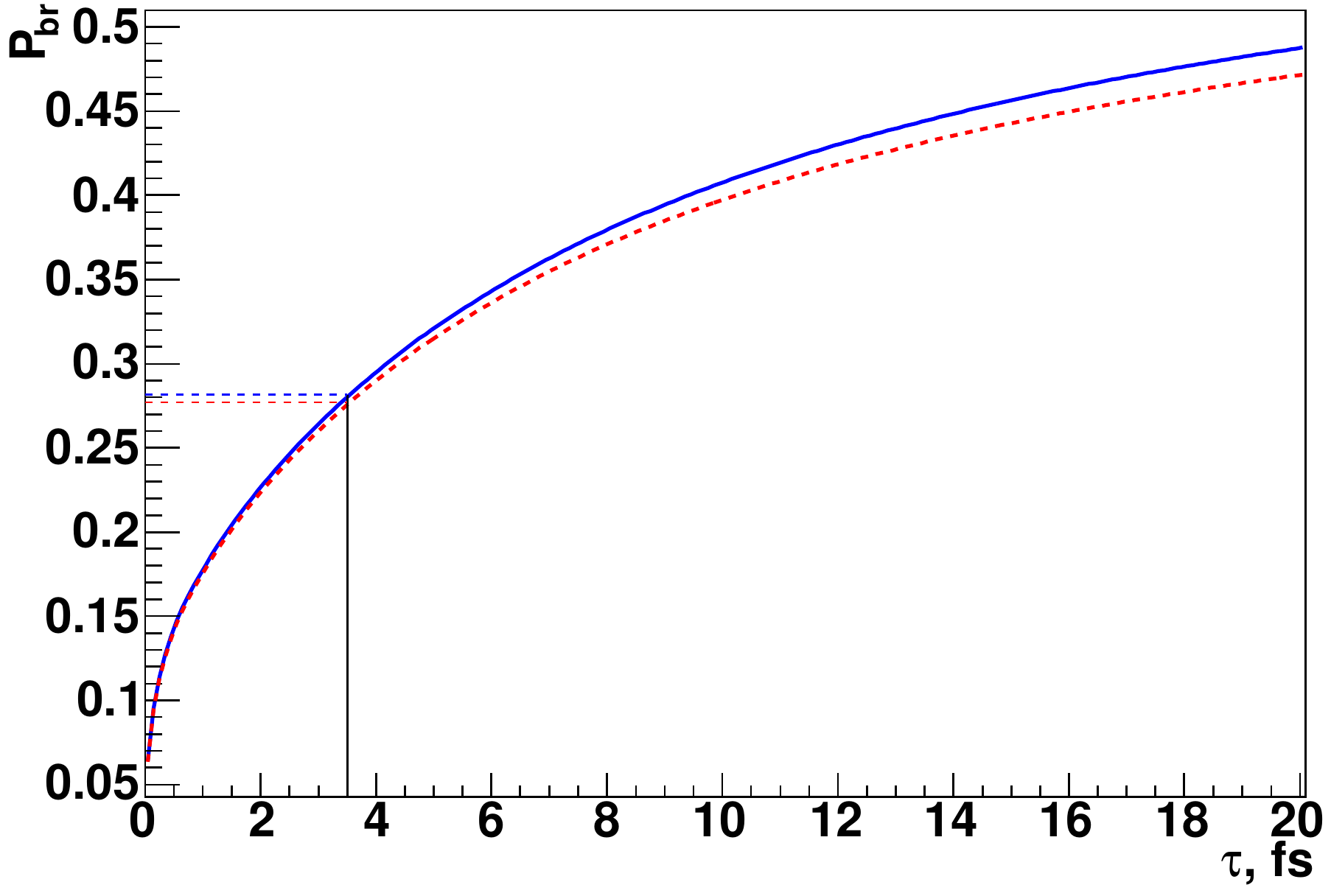}
\caption{Probability of $\pi K$~atom breakup as a function of 
ground state lifetime $\tau$ in Ni targets of thicknesses 
$98\:\mu$m (Ni-1: dashed red) and $108\:\mu$m (Ni-2: solid blue). 
The predicted lifetime $\tau = 3.5\cdot10^{-15}$~s~(Eq.~\ref{eq:julia1}) 
corresponds to the breakup probability $P_{\text{br}}$ = 0.28.}
\label{Fig4_1}
\end{center}
\end{figure}

In the present paper we use a set of total and transition cross sections  
calculated in the first Born approximation
for $\pi K$~atoms interacting with Ni atoms, 
according to the method described in~\cite{Afan96}.
Solving the equation system, the breakup probability  
$P_{\text{br}}(\tau)$ (Fig.~\ref{Fig4_1}) is obtained by convoluting 
$P_{\text{br}}(\tau,p_A)$ with the experimental lab momentum spectra 
of small relative momentum $\pi K$ Coulomb pairs. The function 
$P_{\text{br}}(\tau)$ is used to extract a lifetime 
estimate from the measured $\pi K$~atom breakup probability. 

\section{Data processing}
\label{sec:Data pro}

Recorded events have been reconstructed with  
the DIRAC $\pi\pi$~\cite{ADEV11} analysis software 
ARIANE \cite{Ariane} modified for analysing $\pi K$ data.

\subsection{Tracking and setup tuning}
\label{ssec:Tracking}

Only events with one or two particle tracks 
in the DC of each arm are processed. Event reconstruction is 
performed according to the following steps:
1)~One or two hadron tracks are identified in DC of each arm with hits  
in VH, HH and PSh slabs and no signal in ChN and Mu (Fig.~\ref{Fig2_1}). 
The earliest track in each arm is used for further analysis.
2)~So-called DC tracks are extrapolated backward to  
the incident proton beam position on the target, 
using the transfer function of the DIRAC dipole 
magnet \cite{note0904}. This procedure provides 
approximated particle momenta and  
corresponding intersection points in MDC, SFD and IH.
3)~Hits are searched around the expected SFD coordinates 
in the region defined by position accuracy. 
For events with low and medium background,  
the number of hits around the two tracks is 
$\le 4$ in each SFD plane and  $\le 9$ in all 3 SFD planes. 
These criteria reduce the data sample by 1/3.     
In order to find the best two-track combination, the momentum of 
the positive or negative particle may be modified to match the 
$X$-coordinates of tracks in DC and the SFD hits in the $X$- or $U$-plane. 
Furthermore, the two tracks may not use a common SFD hit 
in case of more than one hit in the proper region. 
In the final analysis the combination with the best $\chi^2$  
in the other SFD planes is kept.

\begin{figure}[h] \centering{
\includegraphics[scale=0.5]{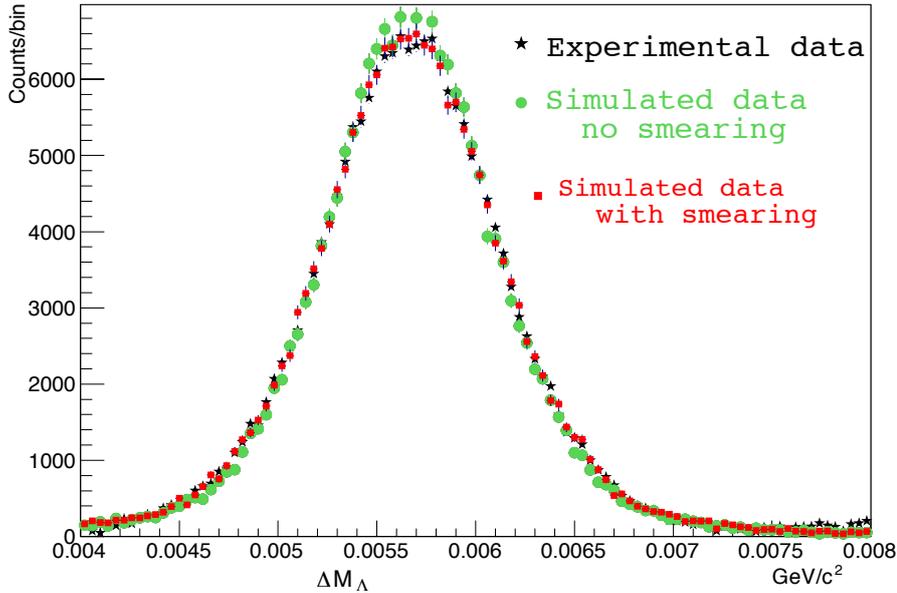}
}
\caption{
Invariant $\pi^-{\rm p}$ mass distribution in the $\Lambda$ region. 
[$\Delta{\rm M}_\Lambda = {\rm M}_\Lambda - 1.11$~GeV/c$^2$; 
green: MC distribution without smearing; 
red: MC with smearing of $7\cdot10^{-4}$;  
black: experimental data] 
}
\label{Fig5_1}
\end{figure}

To check and align the setup components, 
we take advantage of the $\Lambda \to \pi^-{\rm p}$ and 
$\bar{\Lambda} \to \pi^+\bar{\rm p}$ 
decays~\cite{note09og,note0516}.
Using data from 2008 to 2010 and after geometrical alignment, 
the reconstructed $\Lambda$ mass 
[$(1.115685\pm1.2\cdot10^{-6})$~GeV/c$^2$] 
agrees well with the PDG value  
[$(1.115683\pm6\cdot10^{-6})$~GeV/c$^2$]~\cite{note1303,pdg}.  
The width of the $\Lambda$ peak is a tool 
to evaluate the momentum resolution: 
it depends mainly on multiple scattering in the upstream setup part  
and in the Al membrane at the exit of 
the vacuum chamber as well as on DC resolution and alignment.  
The upstream multiple scattering has been 
determined by analysing $\pi \pi$ events \cite{note1204}.  
The MC simulation underestimates the $\Lambda$ width by 6 -- 7\% 
with respect to the experimental value, and this difference is  
consistent for each momentum bin and for $\Lambda$ and 
$\overline{\Lambda}$. Hence we attribute the discrepancy 
between experiment and simulation to an imperfect description of 
the downstream setup part. To fix it, a Gaussian smearing of 
the reconstructed momenta is introduced.  
The smearing applied event-by-event is given by the formula: 
$p^{smeared} = p \; (1 + C_f \cdot N(0,1))$, 
where $p$ is the reconstructed proton or pion momentum and 
$N(0,1)$ a random number generated according to 
the standard normal distribution.  
Smearing of simulated momenta with $C_f=(7\pm4)\cdot10^{-4}$ leads to 
a $\Lambda$ width in the reconstructed MC events consistent with 
experimental data~\cite{note1303} (Fig.~\ref{Fig5_1}).
Using the decays $\Lambda \to \pi^-{\rm p}$ and 
$\bar{\Lambda} \to \pi^+\bar{\rm p}$ and 
taking into account momentum smearing,  
the momentum resolution has been evaluated as  
$ \frac{dp}{p} = \frac{p_{gen} -p_{rec} } {p_{gen}}$  
with $p_{gen}$ and $p_{rec}$ the generated and reconstructed momenta, 
respectively.     
Between 1.5 and 8~GeV/c DIRAC is able to reconstruct particle momenta 
with a relative precision from $2.8\cdot10^{-3}$ to $4.4\cdot 10^{-3}$. 
The following resolutions in ($Q_{X}$, $Q_{Y}$, $Q_{L}$) after 
the target are obtained by MC simulation: 
$\sigma_{QX} \approx \sigma_{QY} \approx 0.18~\rm{MeV}/c$, 
$\sigma_{QL} \approx 0.85~{\rm MeV}/c$ for 
$p_{\pi K} = p_{\pi}+p_K = 5$~GeV/c and about 6\% higher values  
for $p_{\pi K} = 7.5$~GeV/c.

\subsection{Event selection}
\label{ssec:Ev_sel}

\begin{figure}[htbp]
	\begin{center}
		\begin{minipage}[t]{0.45\linewidth}
			\centering
			\includegraphics[width=\linewidth]{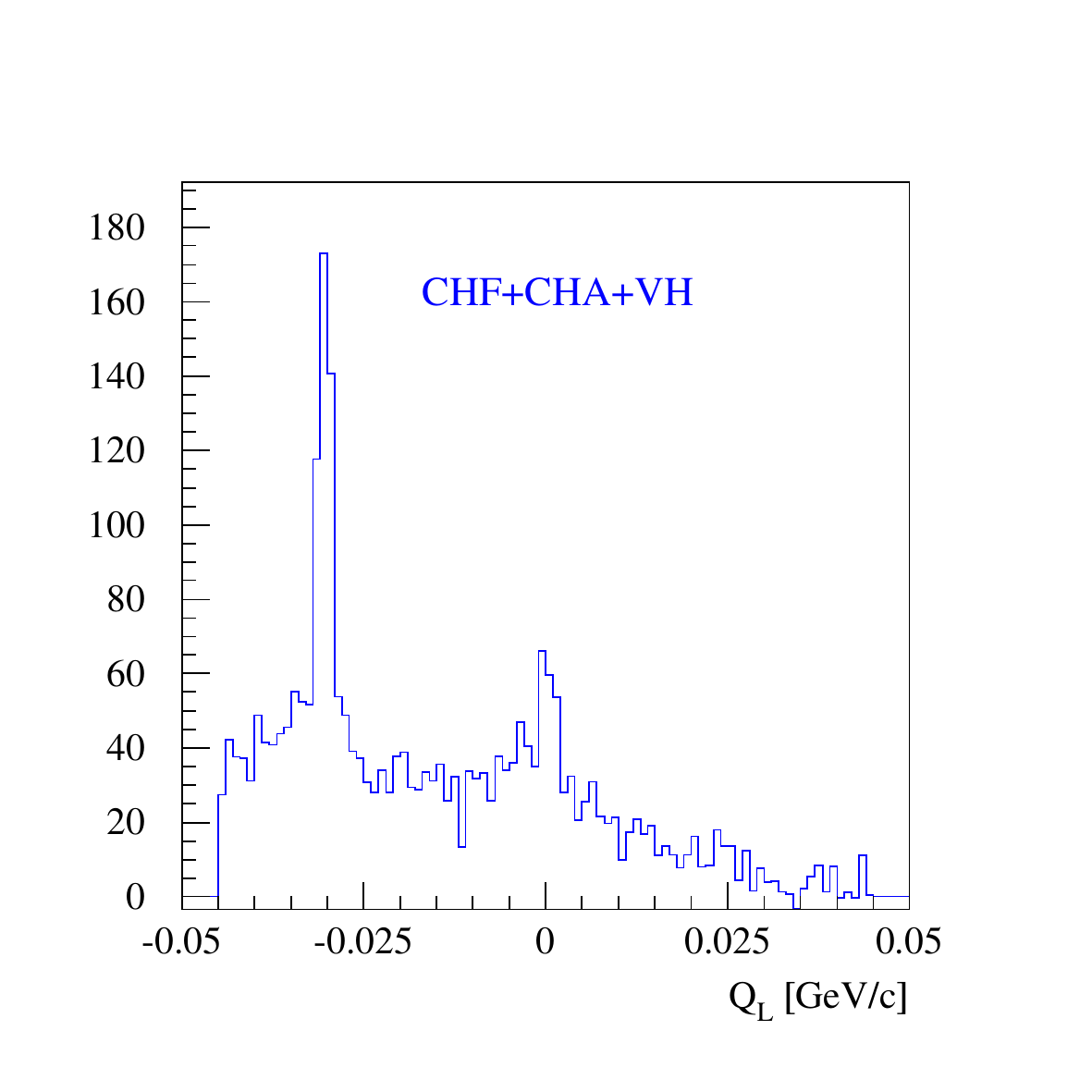}
			\caption{
\footnotesize $Q_L$ distribution of hypothesised $\pi^-K^+$ pairs 
after applying the selection described in the text. Events with 
positive $Q_L$ are suppressed compared to those with negative $Q_L$  
due to lower acceptance and lower production cross section.			
			}
			\label{Fig5_2}
		\end{minipage}
		\qquad
		\begin{minipage}[t]{0.45\linewidth}
			\centering
			\includegraphics[width=\linewidth]{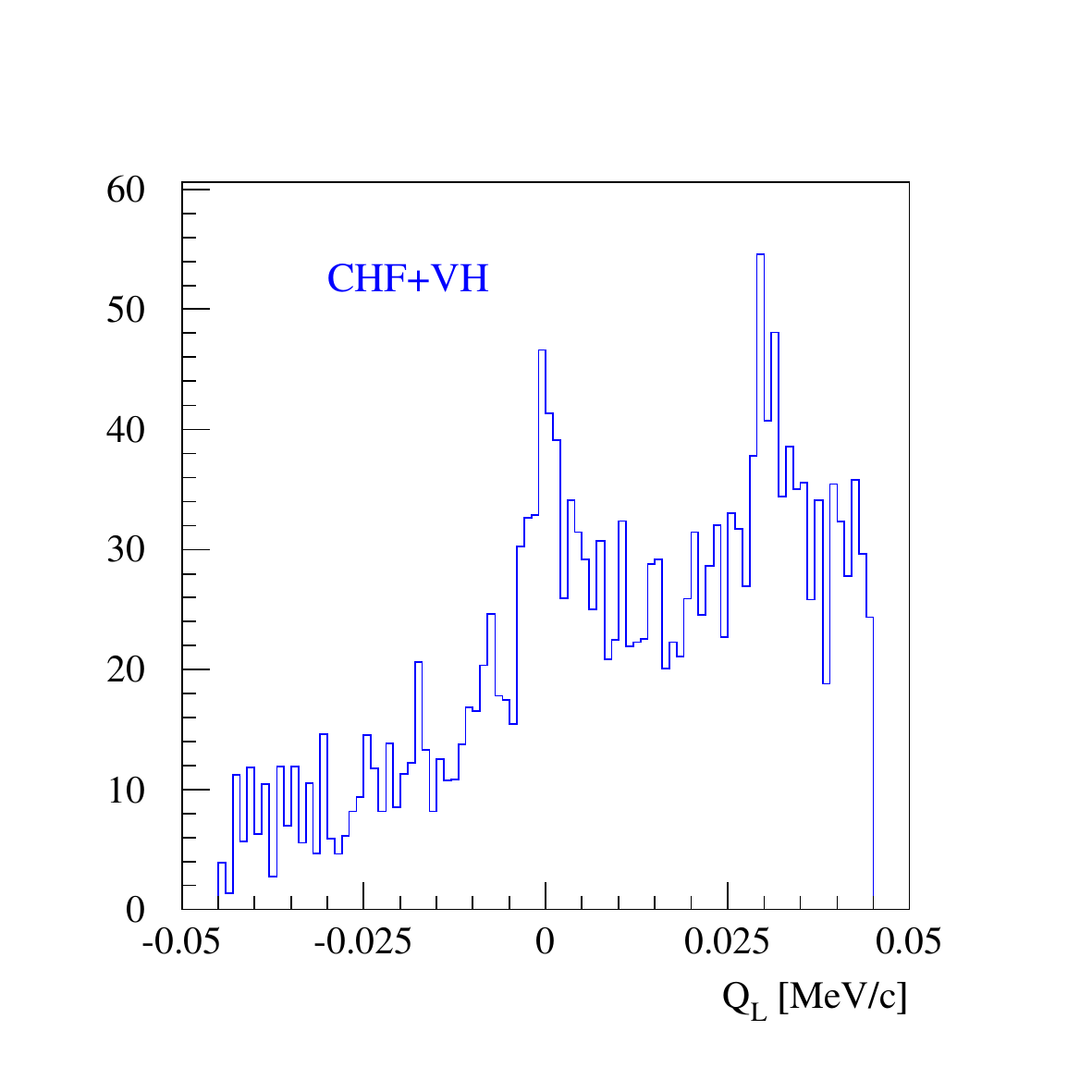}
			\caption{
\footnotesize $Q_L$ distribution of hypothesised $\pi^+K^-$ pairs 
after selection. Events with negative $Q_L$ are suppressed compared to 
those with positive $Q_L$ due to acceptance and cross section.			
			}
			\label{Fig5_3}
		\end{minipage}
	\end{center}
\end{figure}

Selected events are divided into the categories 
$\pi^-K^+$, $\pi^+K^-$ and $\pi^+\pi^-$. 
The last event type is used for calibration purposes. 
Pairs of $\pi K$ are cleaned from  
$\pi^+\pi^-$, $\pi^-{\rm p}$ and $\pi^+\bar{\rm p}$ background by   
the Cherenkov counters ChF and ChA. 
In the momentum range from 3.8 to 7~GeV/c 
pions are detected by ChF with (95 -- 97)\% efficiency \cite{note1305}, 
whereas kaons and protons (antiprotons) do not produce a signal.
The admixture of $\pi^-{\rm p}$ pairs is suppressed by the 
aerogel Cherenkov detector (ChA), which records kaons  
but not protons \cite{note0907}.
By requiring a signal in ChA and 
selecting compatible time-of-flights between target and VH, 
$\pi^-{\rm p}$ and $\pi^-\pi^+$ pairs, contaminating $\pi^-K^+$,  
can be substantially suppressed.
Fig.~\ref{Fig5_2} shows the well-defined $\pi^-K^+$ Coulomb peak 
at $Q_L=0$ and the strongly suppressed peak from $\Lambda$ decays 
at $Q_L=-30$~MeV/c. Similarly Fig.~\ref{Fig5_3} presents 
the $\pi^+K^-$ Coulomb peak at $Q_L=0$ and a second weaker peak 
from $\overline{\Lambda}$ decay at $Q_L=30$ MeV/c~\footnote{
Note that $Q_L(\pi^+ K^-) = -Q_L(\pi^- K^+)$ for the same $p_K/p_{\pi}$.
}. 

The final analysis sample contains only events which fulfil the following criteria:
\begin{equation}\label{eq:7c_critq}
|Q_X| < 6~{\rm MeV/c} \, , \, |Q_Y| < 4~{\rm MeV/c} \, , \, |Q_L| < 15~{\rm MeV/c} \, .
\end{equation}

Due to finite detector efficiency still a certain admixture of 
misidentified pairs remains in the experimental distribution. 
Their contribution has been estimated by time-of-flight investigations 
and accordingly subtracted \cite{note1306}.

\section{Data simulation}
\label{sec:Data sim}

Since the $\pi K$ data samples consist of Coulomb, non-Coulomb and 
atomic pairs, three event types have been generated by MC 
in adequate high statistics. These events are characterised by 
different $q$ distributions: 
the non-Coulomb pairs are uniformly distributed in low $q$, 
while the $q$ distribution for Coulomb pairs is modified by 
the factor $A_C(q)$ (Eq.~\ref{eq:cross_sect_C}).
For each atomic pair one needs to know the position of the breakup and the 
lab momentum. In practice the MC lab momentum distributions  
are approximated by analytic formulae, 
which resemble the experimental momentum distributions 
of such pairs \cite{note1001,note0711}.
After comparing experimental momentum spectra \cite{note1306} with 
MC distributions reconstructed by the analysis software,  
the simulated distributions have been fitted 
to the experimental data by a weight function. 
The breakup point and the quantum numbers of 
the atomic state, from which ionisation occurred, 
are obtained by solving numerically 
the transport equations~\cite{Zhab08}, 
using total and transition cross sections~\cite{Afan96}.
The lab momenta of the atoms are assumed to be the same as 
for Coulomb pairs. The description of the charged particle propagation 
through the setup takes into account: 
a) multiple scattering in the target, detector planes and 
partitions, b) response of all detectors, c) additional smearing 
of particle momentum, d) results of SFD response analysis 
\cite{Gorin06,sfd,note1306} with influence on the $Q_T$ resolution.

\section{Data analysis}
\label{sec:Analysis}

\begin{figure}[!htbp]
\begin{center}
\includegraphics[width=75mm]{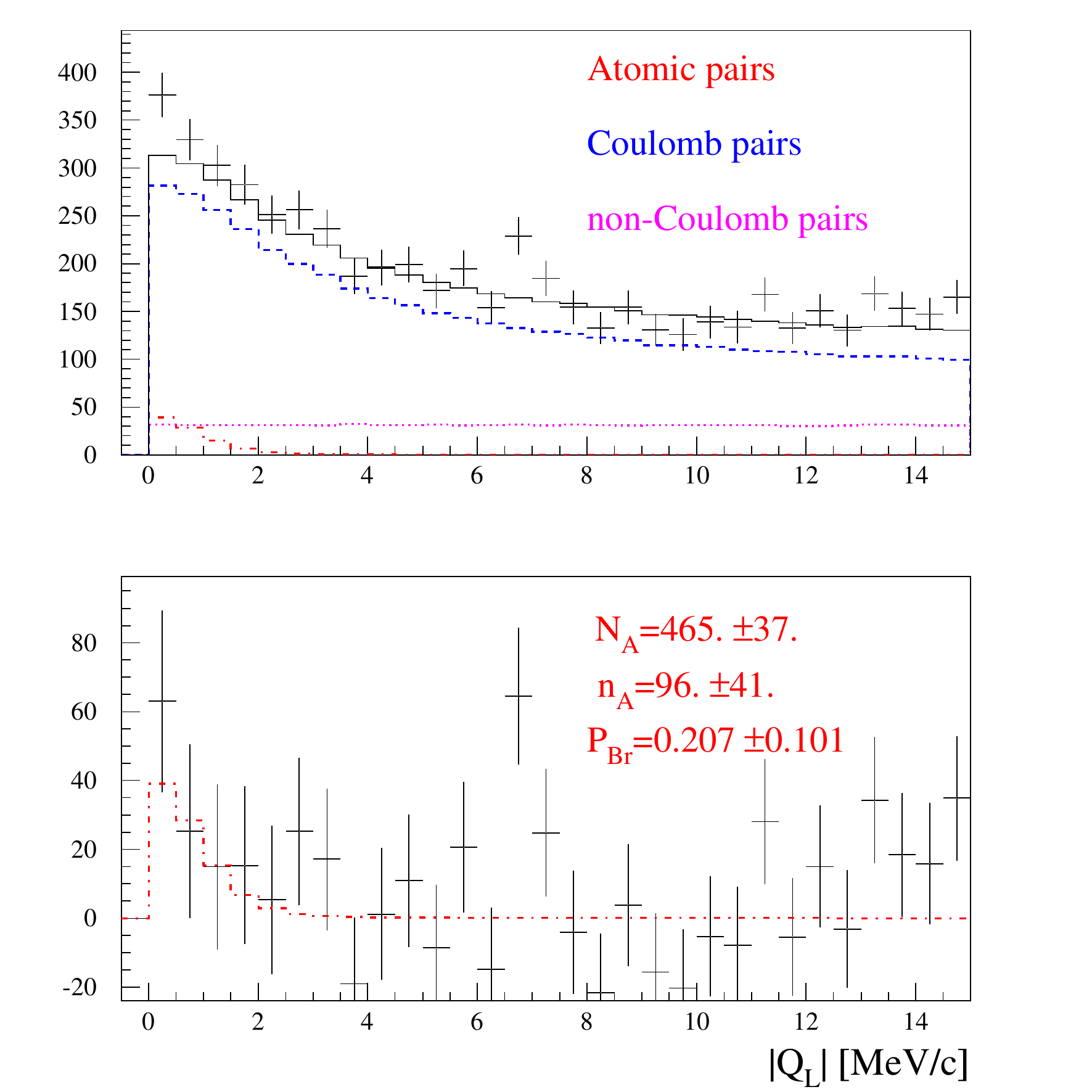}
\caption{\footnotesize Top: Experimental $|Q_L|$ distribution of 
$\pi^-K^+$ pairs [2-dimensional ($Q_T,Q_L$) analysis] fitted by 
the sum of simulated distributions of atomic, Coulomb and 
non-Coulomb pairs. Atomic pairs are shown in red, and 
free pairs (Coulomb and non-Coulomb) in black. 
Bottom: Difference distribution between experimental and 
simulated free pair distributions compared with 
simulated atomic pairs.}
\vspace{-5mm}
\label{Fig7_1}
\end{center}
\end{figure}

\begin{figure}[!htbp]
\begin{center}
\includegraphics[width=75mm]{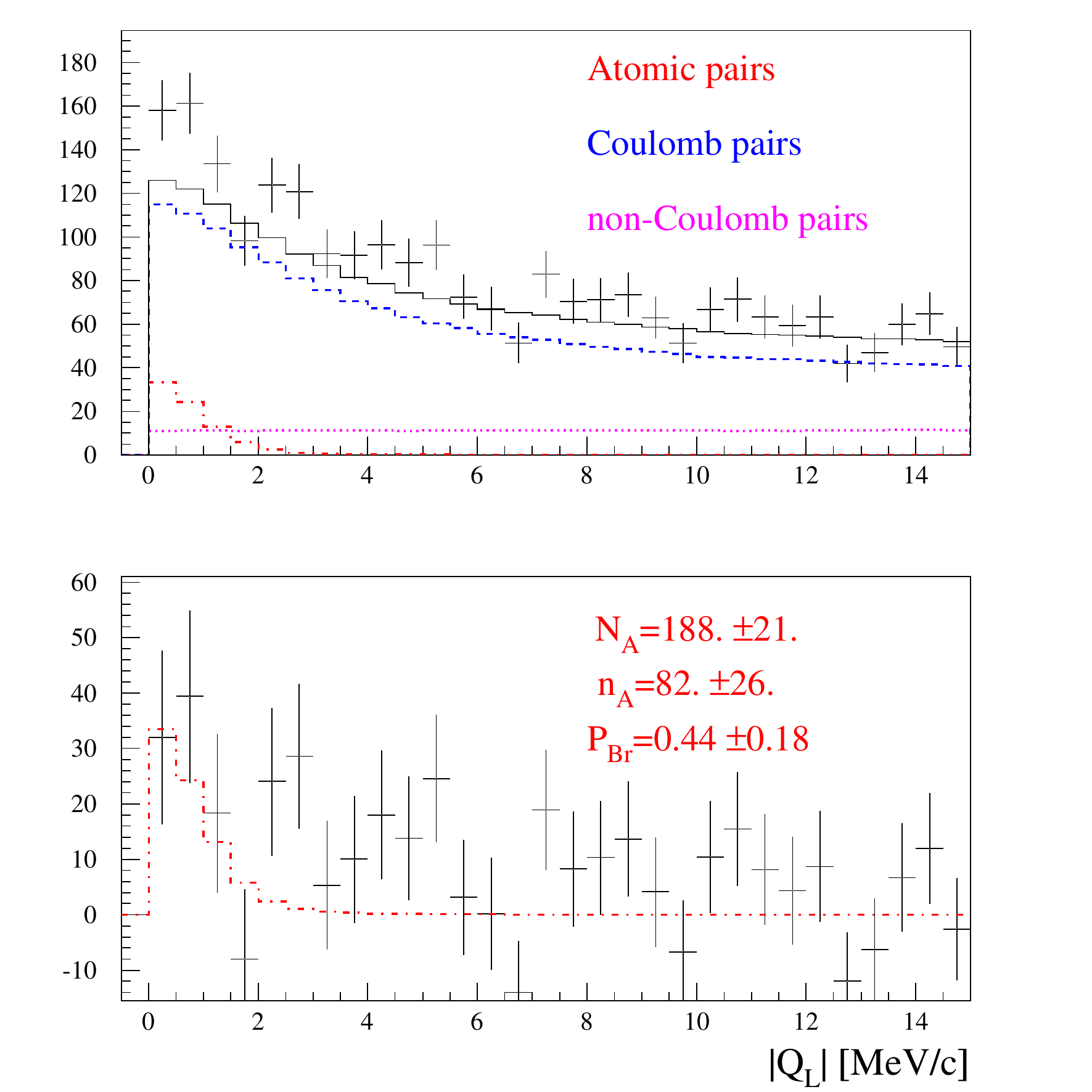}
\caption{\footnotesize Experimental $|Q_L|$ distributions for $\pi^+K^-$ pairs  
analogous to Fig.~\ref{Fig7_1}.}
\vspace{-5mm}
\label{Fig7_2}
\end{center}
\end{figure}

The analysis of $\pi K$ data is similar to that of $\pi^+\pi^-$ 
data~\cite{ADEV11}: experimental distributions of 
relative momentum $Q$ components have been fitted by 
simulated distributions of atomic, Coulomb and non-Coulomb pairs. 
Their corresponding numbers $n_A$, $N_C$ and $N_{nC}$ are 
free fit parameters. The relation (\ref{eq:number_A}) between 
the numbers of produced atoms and Coulomb pairs allows 
to derive the breakup probability. The same procedure has been applied  
to $\pi^-K^+$ (Fig.~\ref{Fig7_1}) and $\pi^+K^-$ (Fig.~\ref{Fig7_2}) pairs. 
The $Q_L$ distributions shown are obtained from 
the 2-dimensional ($Q_T,Q_L$) distributions in the region 
$Q_T < 4~$~MeV/c, $|Q_L| < 15$~MeV/c for pairs with lab momenta 
$4.8<p_{\pi^-}+p_{K^+}<7.2$~GeV/c and 
$4.8<p_{\pi^+}+p_{K^-}<7.6$~GeV/c.  
The different background conditions are taken into account.
One observes an excess of events in Fig.~\ref{Fig7_1} and 
\ref{Fig7_2} in the low $Q_L$ region, where atomic pairs are expected.

Similarly the analysis has been performed for the 1-dimensional ($Q_L$) 
distributions with the results shown in Table~\ref{tab:7_1}. 
The 1- and 2-dimensional distributions have different 
sensitivities to sources of systematic errors \cite{note0804}. 
Comparing the two outcomes allows to check the stability of 
our analysis procedure. The experimental conditions vary 
from 2008 to 2010 due to setup updates and beam quality. 
Table~\ref{tab:7_1} summarises all the fit results of 
the data samples analysed on the basis of the 
2-dimensional as well as the 1-dimensional distributions. 
The number of reconstructed atomic pairs of both charge combinations 
from the 2-dimensional analysis amounts to   
$n_A(\pi^-K^+ + \pi^+K^-) = 178\pm 49$ (3.6 sigma). 
On the basis of this number the extracted values 
for the breakup probability presented in the last column of 
Table~\ref{tab:7_1} provide a means to estimate the $\pi K$ atom lifetime.
\begin{table}[!htb]
\small{
\caption{Results for $N_A$ (number of produced atoms), 
$n_A$ (number of atomic pairs) and $P_{br}$ (breakup probability) 
by analysing 2-dimensional ($Q_T,Q_L$) 
and 1-dimensional ($Q_L$) distributions.}
\label{tab:7_1}
\begin{center}
\begin{tabular}{|r|r|r|r|}
\hline
 Year & \multicolumn{1}{|c|}{$N_A$} & \multicolumn{1}{|c|}{$n_A$} & \multicolumn{1}{|c|}{$P_{br}$} \\
\hline
\multicolumn{4}{|c|}{$\pi^-K^+$ over $Q_T,Q_L$} \\
\hline
 2008 & $132 \pm 16$ & $14 \pm 19$ & $0.11 \pm 0.15$ \\
 2009 & $169 \pm 24$ & $33 \pm 26$ & $0.20 \pm 0.17$ \\
 2010 & $164 \pm 23$ & $49 \pm 26$ & $0.30 \pm 0.19$ \\
\hline
\multicolumn{4}{|c|}{$\pi^-K^+$ over $Q_L$} \\
\hline
 2008 & $125 \pm 19$ & $ 25 \pm 30$ & $0.20 \pm 0.26$ \\
 2009 & $151 \pm 28$ & $ 54 \pm 42$ & $0.36 \pm 0.33$ \\
 2010 & $155 \pm 28$ & $ 61 \pm 42$ & $0.39 \pm 0.32$ \\
\hline
\multicolumn{4}{|c|}{$\pi^+K^-$ over $Q_T,Q_L$} \\
\hline
 2008 & $ 51 \pm 11$ & $21 \pm 13$ & $0.41 \pm 0.33$ \\
 2009 & $ 77 \pm 13$ & $26 \pm 16$ & $0.34 \pm 0.24$ \\
 2010 & $ 60 \pm 12$ & $35 \pm 16$ & $0.58 \pm 0.36$ \\
\hline
\multicolumn{4}{|c|}{$\pi^+K^-$ over $Q_L$} \\
\hline
 2008 & $ 47 \pm 13$ & $35 \pm 21$ & $ 0.75 \pm 0.62$ \\
 2009 & $ 76 \pm 15$ & $28 \pm 24$ & $ 0.37 \pm 0.37$ \\
 2010 & $ 83 \pm 15$ & $-4 \pm 22$ & $-0.04 \pm 0.26$ \\
\hline
\end{tabular}
\end{center}
}
\end{table}

\section{Systematic errors}
\label{sec:systematic}

The evaluation of the breakup probability $P_{\text{br}}$ is 
affected by several sources of systematic errors~\cite{note1306}. 
Most of them are induced by imperfections in the simulation of 
the different $\pi K$ pairs: atomic, Coulomb, non-Coulomb and 
misidentified pairs. Shape differences of experimental and 
simulated distributions in the fit procedure 
(section~\ref{sec:Analysis}) lead to biases on parameters,
including breakup probability. The influence of error sources 
is different for the ($Q_T,Q_L$) and $Q_L$ analyses.
Table~\ref{tab:8_1} shows systematic errors common  
to $\pi^-K^+$ and $\pi^+K^-$ collected from 2008 to 2010. 
\begin{table}[!htb]
\small{
\caption{Systematic errors in $P_{\text{br}}$ common to 
all data collected from 2008 to 2010.}
\label{tab:8_1}
\begin{center}
\begin{tabular}{|p{5cm}|c|c|}
\hline
\rule{0pt}{12pt}Sources of systematic errors & $\sigma^{syst}_{Q_T,Q_L}$ & $\sigma^{syst}_{Q_L}$ \\[0.3em]
\hline
Uncertainty in $\Lambda$ width correction & 0.005 & 0.0015 \\
 & & \\
Accuracy of SFD simulation & 0.0008 & 0.0003 \\
 & & \\
Correction of Coulomb correlation function on finite size production region & 0.00006 & 0.00006 \\
 & & \\
Uncertainty in $P_{br}(\tau)$ dependence & 0.005 & 0.005 \\
 & & \\
Uncertainty in target thickness & 0.0003 & $<0.0003$ \\
\hline
\end{tabular}
\end{center}
}
\end{table}
Other sources of systematic errors are uncertainties in the  
measuring procedure for $\pi K$ and background distributions. 
These spectra have been measured individually for the different run periods,   
producing systematic errors $\sigma^{syst}_{\pi K}$ and 
$\sigma^{syst}_{back}$ in $P_{br}$ (see Table~\ref{tab:8_2}). 
The presented systematic errors have been included in estimating 
the $\pi K$ atom lifetime as described in the next section.
\begin{table}[!htb]
\small{
\caption{Systematic errors in $P_{\text{br}}$ specific to 
the data samples collected in 2008, 2009 and 2010.}
\label{tab:8_2}
\begin{center}
\begin{tabular}{|c|c|c|}
\hline
\rule{0pt}{12pt}Year & $\sigma^{syst}_{\pi K}$ & $\sigma^{syst}_{back}$ \\[0.3em]
\hline
\multicolumn{3}{|c|}{$K^+\pi^-$ over $Q_T,Q_L$} \\
\hline
 2008 & 0.0028 & 0.0015 \\
 2009 & 0.0044 & 0.0025 \\
 2010 & 0.0036 & 0.0022 \\
\hline
\multicolumn{3}{|c|}{$K^+\pi^-$ over $Q_L$} \\
\hline
 2008 & 0.0030 & 0.0028 \\
 2009 & 0.0053 & 0.0044 \\
 2010 & 0.0046 & 0.0036 \\
\hline
\multicolumn{3}{|c|}{$\pi^+K^-$ over $Q_T,Q_L$} \\
\hline
 2008 & 0.0072 & 0.0067 \\
 2009 & 0.0048 & 0.0028 \\
 2010 & 0.0017 & 0.0043 \\
\hline
\multicolumn{3}{|c|}{$\pi^+K^-$ over $Q_L$} \\
\hline
 2008 & 0.0093 & 0.0072 \\
 2009 & 0.0047 & 0.0048 \\
 2010 & 0.0021 & 0.0017 \\
\hline
\end{tabular}
\end{center}
}
\end{table}

\section{Lifetime and scattering length measurements}
\label{sec:measurement}

The lifetime dependence of the breakup probability  
$P_\text{br}(\tau, p_A)$ for $\pi^{\mp}K^{\pm}$ atoms 
with momentum~$p_A$ has been determined~\cite{Zhab08},  
using total and excitation cross~sections calculated in 
Born approximation~\cite{Afan96}. Convoluting 
$P_\text{br}(\tau, p_A)$ with the corresponding lab    
momentum spectra (section~\ref{sec:Atom_int} and \cite{note1306})  
leads to a set of $P_{\text{br},i}(\tau)$ functions, 
each for every target thickness (Ni-1, Ni-2) and 
experimental spectrum ($\pi^+K^-$, $\pi^-K^+$).
To estimate the ground state lifetime  
the maximum likelihood method~\cite{note0807} 
has been applied: 
\begin{equation}
 L(\tau) = \exp \left(-U^T G^{-1} U/2\right),
\end{equation}
where $U$ with $U_i=\Pi_i-P_{\text{br},i}(\tau)$ is a vector of 
differences between measured~$\Pi_i$ ($P_{br}$ in Table~\ref{tab:7_1}) 
and theoretical breakup probability $P_{\text{br},i}(\tau)$ 
for data sample~$i$. The matrix $G$, the error matrix of $U$, 
includes statistical and systematic uncertainties 
(Table~\ref{tab:8_1} and \ref{tab:8_2}):
\begin{equation}\label{eq:Gij}
 G_{ij} = \delta_{ij}
 \left[(\sigma^{\text{stat}}_i)^2
 + (\sigma^{\text{syst}}_{\pi K,i})^2
 + (\sigma^{\text{syst}}_{\text{back},i})^2
 \right]
 + (\sigma^{\text{syst}}_{\text{global}})^2.
\end{equation}
By combining the two charge combinations ($\pi^{\mp}K^{\pm}$) 
and considering the statistics collected from 2008 to 2010,
the $(Q_T,Q_L)$ analysis yields the following 
ground state lifetime estimation: 
\begin{equation}
 \tau = 
    \left.\left.(2.5_{-1.8}^{+3.0}\right|_{\text{stat}}
                  {}_{-0.1}^{+0.3}\right|_{\text{syst}}\:)\text{fs} =
    \left.(2.5_{-1.8}^{+3.0}\right|_{\text{tot}}\:)\text{fs}.
 \label{eq:tau_QTQL}
\end{equation}
This experimental value agrees with the predicted one 
of Eq.~(\ref{eq:julia1}).

The estimated ground state lifetime~(\ref{eq:tau_QTQL}) 
corresponds to the $\pi K$ scattering length~(\ref{eq:julia0}) 
\begin{equation}\label{eq:a0exp}
 \left|a_0^-\right|M_{\pi}=
 \frac{1}{3}\left|a_{1/2}-a_{3/2}\right|M_{\pi}=
  0.107_{-0.035}^{+0.093} = 0.11_{-0.04}^{+0.09} \:,
\end{equation}
to be compared with the theoretical predictions (\ref{eq:chpt}).

The $Q_L$~analysis (Table~\ref{tab:7_1}, \ref{tab:8_1} and \ref{tab:8_2}) 
provides a similar estimation of the ground state lifetime, 
but with worse precision:
\begin{equation}
 \tau =
    \left.\left.(2.4_{-2.2}^{+5.4}\right|_{\text{stat}}
                  {}_{-0.1}^{+0.5}\right|_{\text{syst}}\:)\text{fs} =
    \left.(2.4_{-2.2}^{+5.5}\right|_{\text{tot}}\:)\text{fs}.
 \label{eq:tau_QL}
\end{equation}

\section{Conclusion}
\label{sec:conclusion}

The analysis of $\pi K$ pairs collected from 2008 
to 2010 allows to evaluate the number of 
atomic $\pi K$ pairs ($178 \pm 49$) as well as 
the number of produced $\pi K$ atoms ($653 \pm 42$) 
and thus the breakup (ionisation) probability. By exploiting 
the dependence of breakup probability on atom lifetime, 
a value for the $\pi K$~atom 1S lifetime 
$\tau=(2.5_{-1.8}^{+3.0})$~fs has been extracted.  
As the atom lifetime is related to a scattering length, 
a measurement of the S-wave isospin-odd $\pi K$~scattering length 
$\left|a_0^-\right|=\left( 0.11_{-0.04}^{+0.09} \right)M_{\pi}^{-1}$ 
can be presented, compatible with theory.

\section*{Acknowledgements}

We are grateful to R.~Steerenberg and the CERN-PS crew for 
the delivery of a high quality proton beam and 
the permanent effort to improve the beam characteristics. 
The project DIRAC has been supported by 
the CERN and JINR administration, 
Ministry of Education and Youth of the Czech Republic by project LG130131, 
the Istituto Nazionale di Fisica Nucleare and the University of Messina (Italy),  
the Grant-in-Aid for Scientific Research from 
the Japan Society for the Promotion of Science, 
the Ministry of Education and Research (Romania), 
the Ministry of Education and Science of the Russian Federation and 
Russian Foundation for Basic Research, 
the Direcci\'{o}n Xeral de Investigaci\'{o}n, Desenvolvemento e Innovaci\'{o}n, 
Xunta de Galicia (Spain) and the Swiss National Science Foundation.

\end{document}